\documentclass[12pt,preprint]{aastex}

\usepackage[usenames]{color}%for preprints

\slugcomment{ICRR-Report 624-2012-13:  IPMU 12-0163}
\shorttitle{Production of $^9$B\lowercase{e} through $\alpha$-fusion reaction}
\shortauthors{Kusakabe \& Kawasaki}

\begin{document}

\title{PRODUCTION OF $^9$B\lowercase{e} THROUGH THE $\alpha$-FUSION REACTION OF METAL-POOR COSMIC RAYS AND STELLAR FLARES}

\author{Motohiko Kusakabe\altaffilmark{1} and Masahiro Kawasaki\altaffilmark{1,2}}

\altaffiltext{1}{Institute for Cosmic Ray Research, University of Tokyo,
  Kashiwa, Chiba 277-8582, Japan\\
{\tt kusakabe@icrr.u-tokyo.ac.jp, kawasaki@icrr.u-tokyo.ac.jp}}
\altaffiltext{2}{Kavli Institute for the Physics and Mathematics of the
  Universe, University of Tokyo, Kashiwa, Chiba 277-8568, Japan\\
}

\begin{abstract}
Spectroscopic observations of metal-poor stars have indicated possible $^6$Li abundances that are much larger than the primordial abundance predicted in the standard big bang nucleosynthesis model.  Possible mechanisms of $^6$Li production in metal-poor stars include pregalactic and cosmological cosmic-ray (CR) nucleosynthesis and nucleosynthesis by flare-accelerated nuclides.  We study $^9$Be production via two-step $\alpha$-fusion reactions of CR or flare-accelerated $^{3,4}$He through $^6$He and $^{6,7}$Li, in pregalactic structure, intergalactic medium, and stellar surfaces.  We solve transfer equations of CR or flare particles and calculate nuclear yields of $^6$He, $^{6,7}$Li, and $^9$Be taking account of probabilities of processing $^6$He and $^{6,7}$Li into $^9$Be via fusions with $\alpha$ particles.  Yield ratios, i.e., $^9$Be/$^6$Li, are then calculated for the CR and flare nucleosynthesis models.  We suggest that the future observations of $^9$Be in metal-poor stars may find enhanced abundances originating from metal-poor CR or flare activities.
\end{abstract}

%nuclear reactions, nucleosynthesis, abundances  --- solar system: general --- supernovae: general

\section{INTRODUCTION}\label{sec1}

$^6$Li/$^7$Li isotopic ratios of metal-poor stars (MPSs) have been measured spectroscopically.  A possible plateau abundance of $^6$Li/H$\sim 6\times10^{-12}$ has been suggested~\citep{asp2006}, which is about 1000 times higher than the prediction of the standard big bang nucleosynthesis (BBN) model.  Such a high abundance of $^6$Li can, however, be derived erroneously because of asymmetries in atomic line profiles originating from convective motions in atmospheres~\citep{cay2007}.  The effect of the convection-driven line asymmetries was recently estimated and it was reported that high $^6$Li abundances have been likely detected in only a reduced number of MPSs \citep[at most several MPSs;][]{asp2008,gar2009,ste2010,ste2012}.   The high abundance level cannot be explained in standard Galactic cosmic-ray (CR) nucleosynthesis models \citep{men1971,ree1974,ram1997,van2000,fie2000,val2002} since the models predict $^6$Li abundances much smaller than the observed level at the low-metallicity region of [Fe/H] $<-2$~\citep{pra2006}.

There are three different classes of astrophysical models for explanations of high $^6$Li abundances in MPSs which assume astrophysical energy sources for nuclear reactions producing $^6$Li.

The first model is the cosmological CR (CCR) nucleosynthesis model, in which $^6$Li is produced via the $\alpha+\alpha$ fusion and spallation of CNO nuclei~\citep{mon1977,rol2005,rol2006,rol2011}.\footnote{We note that Equation (1) of \citet{rol2005}, Equation (11) of \citet{rol2006}, and Equation (2) of \citet{evo2008} are all wrong in the same way.  The CR injection spectrum should be given by a power law in momentum, i.e., Equation (9) of \citet{kus2008}, in order to obtain a consistent formulation (E. Rollinde 2007, private communication).}  \citet{evo2008} have claimed that a CCR model based on a dedicated hierarchical model of Galaxy formation does not reproduce a $^6$Li abundance level or its plateau shape.  The reason for the small $^6$Li abundance is a suppressed star formation rate at high redshift they adopted.\footnote{We note that Equation (8) of \citet{evo2008} for the cross section of $^4$He($\alpha$, $X$)$^6$Li involves a typographical error.  The correct equation is $\sigma_l(E)=66~\exp[-0.0159(4E/{\rm MeV})]$~mb~\citep{mer2001}, with $\sigma_l$ the cross section for the reaction $\alpha+\alpha\rightarrow ^6$Li+$X$, and $E$ the kinetic energy per nucleon of the incident $\alpha$ particle (C. Evoli 2012, private communication).}  Since this CCR model should include CNO spallation, $^9$Be and $^{10,11}$B are necessarily coproduced~\citep{kus2008,rol2008}.  This model assumes CR activities at a typical redshift of $z=\mathcal{O}$(1--10) before Galaxy formation.

The second model is the pregalactic CR (PCR) nucleosynthesis model, in which $^6$Li is produced via the $\alpha+\alpha$ fusion reaction between CR $\alpha$ accelerated in structure formation shocks developed in an early epoch of Galaxy and interstellar $\alpha$ \citep{suz2002}.  This process operates in the epoch of structure formation until the formation of observed stars.

In the above two models, the index of the CR source spectrum with a power law in momentum should be $\gamma \sim 3$ for production of significant amounts of $^6$Li since the smaller and larger indexes fail to predict high $^6$Li abundances as observed in MPSs \citep[see Figure 6 of][]{kus2008}.

The third model is nucleosynthesis by flare-accelerated energetic nuclides, in which $^6$Li is produced mainly via the $^3$He+$\alpha$ reaction between flare-accelerated $^3$He and $\alpha$ in stellar atmospheres \citep{tat2007,tat2008}.  The $^3$He+$\alpha$ reaction has been introduced in a study on $^6$Li production in solar flares \citep{ram2000}.  Flare nucleosynthesis enhances $^6$Li in stellar atmospheres from the time of star formation to the present.  \citet{tat2007} calculated the nucleosynthesis assuming that the source energy spectrum of flare-accelerated particles is an unbroken power law in kinetic energy of spectral index $s=4\pm 1$~\citep{ram1996}, and that the number ratio $^3$He/$^4$He of accelerated particles is 0.5.

Another possibility for $^6$Li production in the early universe is the $\alpha+\alpha$ fusion reaction~\citep{nak2006} associated with simultaneous CNO spallation~\citep{fie1996,fie2002,nak2004} in Type Ic supernova (SN) explosions through reactions between SN ejecta and interstellar matter (ISM) including circumstellar matter (CSM).  Type Ic SNe possibly contribute to abundances of $^6$Li, Be, and B, in the part of MPSs that are composed of material ejected from the SNe, while the three mechanisms described above somewhat homogeneously enhance the abundances.  The light element production in this model is similar to that in the PCR model except for differences in energy spectra and $^4$He abundances of projectile and target matters.  We show that this model can also produce $^9$Be via the two-step $\alpha$-fusion reactions in this paper.

The nuclide $^9$Be is thought to be produced in the Galaxy predominantly through spallation of carbon, nitrogen, and oxygen at reactions with protons and $^4$He~\citep[][see \citet{pra2012} for a recent comprehensive theoretical study in light of astronomical observations]{ree1970,men1971,ree1974}.  Abundances of $^9$Be in many MPSs have been measured.  The $^9$Be abundance increases nearly in proportion to Fe abundance~\citep{boe1999,pri2000a,pri2000b,pri2002,boe2006,tan2009,smi2009,ito2009,ric2009,boe2011}.  The severest lower limit on the primordial Be abundance, i.e., log(Be/H)$<-14$, has been deduced from an observation of the carbon-enhanced MPS BD+44$^\circ$493 with an iron abundance [Fe/H]$=-3.7$ with Subaru/HDS \citep{ito2009}.  The relation between abundances of Be and iron and also B~\citep{dun1997,gar1998,pri1999,cun2000} and iron is explained by Galactic CR (GCR) nucleosynthesis models including primary and secondary reactions between CNO nuclides and $p$ and $\alpha$~\citep[e.g.,][]{ram1997,van2000,fie2000,val2002,pra2012}.

As for the primordial abundance of $^9$Be, a very small abundance of $^9$Be is produced in BBN at the cosmic time of $t\lesssim 200$ s \citep{coc2012}.  Although the $^9$Be production operates in BBN through reactions including $^6$Li($\alpha$, $p$)$^9$Be, a thermal condition of BBN never produces observationally significant amounts of $^9$Be.  If nonthermal nuclides existed in an epoch after BBN, however, $^9$Be can be generated through $\alpha$-fusion reactions by nonthermal $^6$He and $^6$Li nuclei which are produced via $\alpha+\alpha$ fusion reactions \citep{pos2011}.  In nonthermal nucleosynthesis triggered by CRs or stellar flares, $^9$Be production by two-step nonthermal reactions also occurs.  This $^9$Be production has, however, not been taken into account in the three models for $^6$Li production in the early epoch mentioned above.

Therefore, in this paper we calculate production rates of $^9$Be as well as $^6$He and $^{6,7}$Li in the PCR, CCR, and flare nucleosynthesis models.  The structure of this paper is as follows.  The nuclear reactions that we study are explained in Section~\ref{sec2}.  Our nucleosynthesis model for source energy spectra, nuclear transfers, reaction yields, and input parameters are described in Section~\ref{sec3}.  Results of nuclear yields are shown in Section~\ref{sec4}, and predicted $^9$Be/$^6$Li ratios are presented in Section \ref{sec5}.  We summarize this study in Section~\ref{sec6}.

\section{NUCLEAR REACTIONS}\label{sec2}

The $\alpha+\alpha$ fusion reaction between energetic $\alpha$ and background $\alpha$ produces $^6$Li and $^7$Li \citep{ree1970,men1971,pag1997}.  $^6$Li production via $^3$He($\alpha$, $p$)$^6$Li can also be important if energetic $^3$He nuclei exist abundantly as in a solar flare \citep{tat2007}.  In this paper, we study $^9$Be production via the $\alpha$-fusion of $^6A$ and $^7A$ nuclei produced via the reactions $\alpha+\alpha$ and  $^3$He$+\alpha$.  This production proceeds mainly through the following reactions:
\begin{eqnarray}
^6{\rm He}(\alpha, n)^9{\rm Be},\label{eq1}\\
^6{\rm Li}(\alpha, p)^9{\rm Be},\label{eq2}\\
^7{\rm Li}(\alpha, d)^9{\rm Be}.\label{eq3}
\end{eqnarray}

Energetic nuclei of astrophysical origins usually have distribution functions that include a large number of low-energy particles and a small number of high-energy particles.  CR nucleosynthesis is then dominantly contributed by nuclear reactions at low energies.  At low energies, cross sections of two-body final states tend to be larger than those of three or more bodies \citep{mey1972,rea1984}.  We, therefore, assume that contributions from $^7$Li($\alpha$, $2p$)$^9$Li, $^7$Li($\alpha$, $pn$)$^9$Be, and $^7$Be($\alpha$, $2p$)$^9$Be are smaller than those from $^7$Li($\alpha$, $d$)$^9$Be (Equation (\ref{eq3})).  We then consider only the last reaction, whose cross section has been measured.  We did not find data for $^6$He($\alpha$, $p$)$^9$Li (and its inverse reaction).  This reaction is neglected since the threshold is very high (12.2233 MeV).  

\begin{figure}[htbp]
\begin{center}
\includegraphics[scale=0.4]{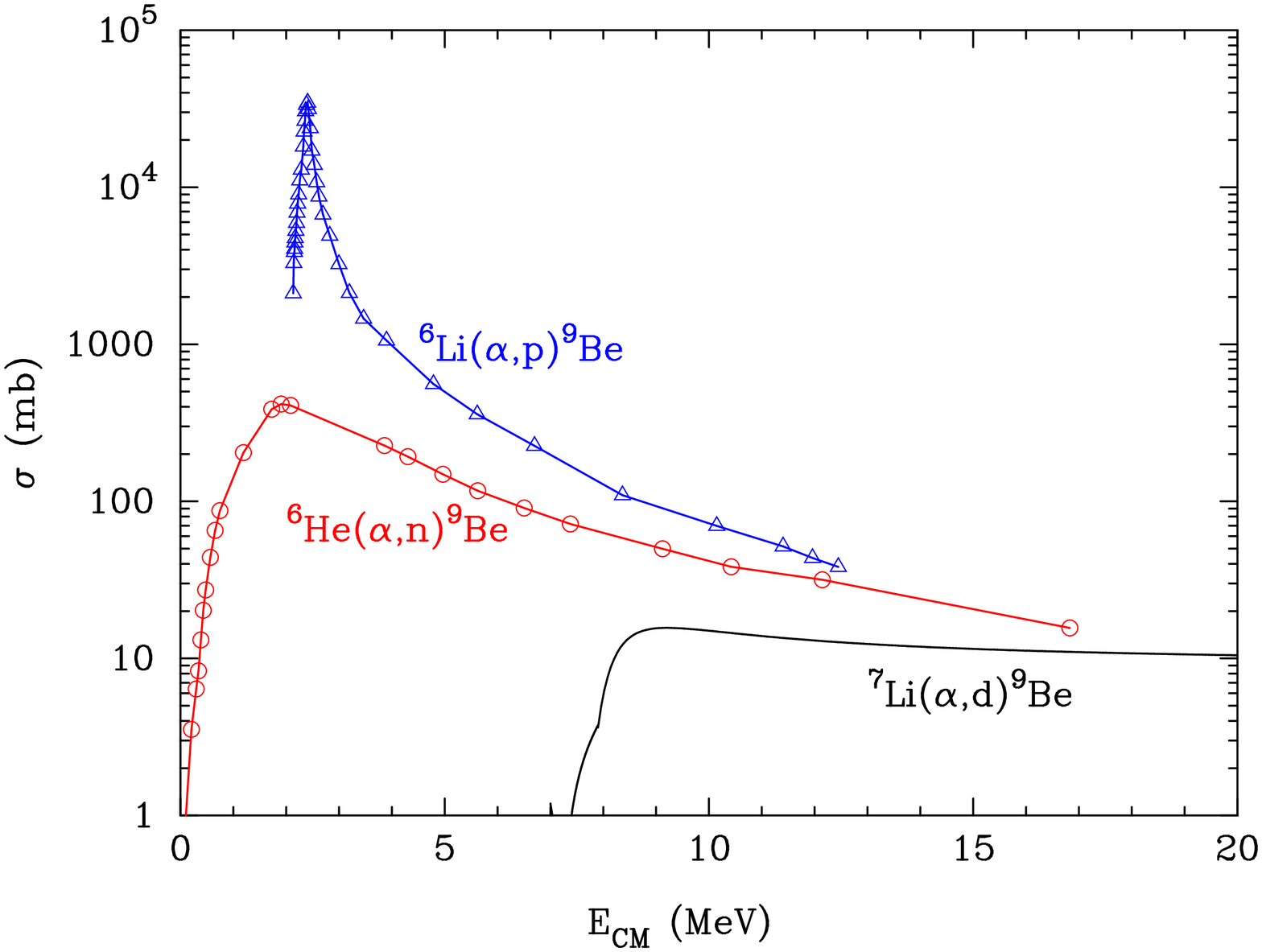}
\end{center}
\caption{Adopted cross sections as a function of center of mass kinetic energy $E_{\rm CM}$ for reactions $^6$He($\alpha$, $n$)$^9$Be \citep{cha2011}, $^6$Li($\alpha$, $p$)$^9$Be \citep{ang1999}, and $^7$Li($\alpha$, $d$)$^9$Be (this study).
\label{fig1}}
\end{figure}

Figure \ref{fig1} shows cross sections as a function of center of mass (CM) kinetic energy, i.e., $E_{\rm CM}$, of $^6$He($\alpha$, $n$)$^9$Be from the Evaluated Nuclear Data File (ENDF/B-VII.1, 2011)~\citep{cha2011}, $^6$Li($\alpha$, $p$)$^9$Be from \citet{ang1999}, and $^7$Li($\alpha$, $d$)$^9$Be estimated in the following subsection.  Since most of the available cross section data are for inverse reactions, we derive forward cross sections for the three reactions utilizing the detailed balance relation~\citep{pag1997}.  

%\placefigure{fig1}

Tables \ref{tab1} and \ref{tab2} show the nuclear mass and spin of species involved in reactions we consider, and reaction $Q$ values, respectively.

\begin{deluxetable}{ccc}
\tablecaption{Mass and Spin of Nuclide
 \label{tab1}}
\tablewidth{0pt}
\tablehead{
\colhead{Species}   & \colhead{Nuclear Mass (GeV)\tablenotemark{a}} & \colhead{Spin}}
 \startdata
$n$       &  0.93956  &  1/2$^+$  \\
$p$       &  0.93827  &  1/2$^+$  \\
$^2$H     &  1.87561  &  1$^+$    \\
$^4$He    &  3.72737  &  0$^+$    \\
$^6$He    &  5.60553  &  0$^+$    \\
$^6$Li    &  5.60151  &  1$^+$    \\ 
$^7$Li    &  6.53383  &  3/2$^-$  \\
$^9$Be    &  8.39274  &  3/2$^-$  \\
\enddata
\tablenotetext{a}{Derived with atomic mass data from \citet{aud2003}.}
\end{deluxetable}

\begin{deluxetable}{cc}
\tablecaption{$Q$ Values Calculated with Mass Data
 \label{tab2}}
\tablewidth{0pt}
\tablehead{
\colhead{Reaction}   & \colhead{$Q$ Value (MeV)}}
 \startdata
$^6$He($\alpha$, $n$)$^9$Be    &  $0.601$  \\
$^6$Li($\alpha$, $p$)$^9$Be    & $-2.124$  \\
$^7$Li($\alpha$, $d$)$^9$Be    & $-7.150$  \\
\enddata
\end{deluxetable}

%\placetable{tab1}
%\placetable{tab2}

We take cross sections of $^4$He($\alpha$, $2p$)$^6$He, $^4$He($\alpha$, $X$)$^6$Li, $^4$He($\alpha$, $p$)$^7$Li, and $^4$He($\alpha$, $n$)$^7$Be from \citet{rea1984} for the kinetic energy of $\alpha$ in the laboratory system $E_\alpha\leq 60$ MeV, and from \citet{mer2001} for $E_\alpha> 60$ MeV.  In addition, we adopt the cross section of $^3$He($\alpha$, $p$)$^6$Li from \citet{cyb2003}.

\subsection{$^7$Li($\alpha$, $d$)$^9$Be Cross Section}\label{sec21}

\begin{figure}[htbp]
\begin{center}
\includegraphics[scale=0.4]{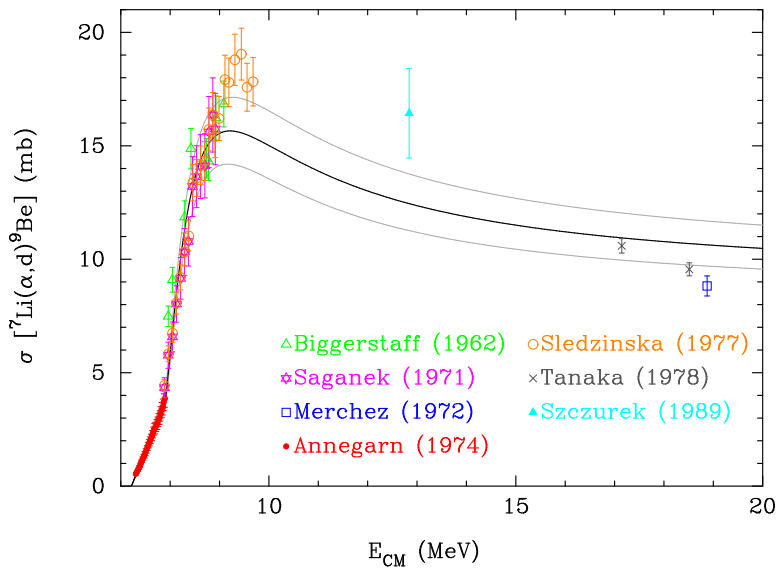}
\end{center}
\caption{Cross section data for the reaction $^7$Li($\alpha$, $d$)$^9$Be as a function of center of mass kinetic energy $E_{\rm CM}$.  The two thick lines are fitting functions in $E_{\rm CM}< 7.9$ MeV and $E_{\rm CM}\geq 7.9$ MeV, respectively.  The thin lines enclose cross section ranges predicted with fitting parameters within $1\sigma$ ranges.
\label{fig2}}
\end{figure}

Figure \ref{fig2} shows cross section data as a function of $E_{\rm CM}$.  We adopt the following data: \citet[][inverse reaction, total of 6 \% errors are adopted]{big1962}, \citet[][inverse, 20 \% errors]{yan1964}, \citet[][inverse, 10 \% errors]{sag1971}, \citet[][forward, typically 5 \% errors as read from their Figure 4]{mer1972}, \citet[][inverse, uncertainties are not published, and we assume 10 \% errors]{ann1974}, \citet[][inverse, 6 \% errors]{sle1977}, \citet[][inverse, 3\% errors]{tan1978}, \citet[][inverse; integration of their Figure 5, 12 \% errors]{szc1989}.

%\placefigure{fig2}

The data are fitted with two functions since the excitation curve at $E_{\rm CM} <7.9$ MeV shows a somewhat shallow slope.  The data in the region $E_{\rm CM} <7.9$ MeV are fitted with a line with an intercept, i.e., $\sigma(E_{\rm CM})=a x+b$, with $x=E_{\rm CM}-E_{\rm CM}^{\rm th}$, where $E_{\rm CM}^{\rm th}=-Q$ is the energy threshold in the CM system.  The best-fit parameter set is $a=5.50$ mb MeV$^{-1}$ and $b=-0.365$ mb.  The fit is good, and the chi-square value is 15.4 for the number of degrees of freedom $30-2=28$.  We then adopt this fitted cross section when its value is positive, while we set the cross section value to be zero otherwise.  The data in the region $E_{\rm CM} \geq 7.9$ MeV are, on the other hand, fitted with the extended Freundlich model, i.e., $\sigma(E_{\rm CM})=c (x/{\rm MeV})^{d (x/{\rm MeV})^{-e}}$.  The best-fit parameter set is $c=8.96$ mb, $d=2.1$, and $e=1.384$.  The fit is rather poor, and the chi-square value is 153 for the number of degrees of freedom $45-3=42$.  In Figure \ref{fig2}, best-fit curves are drawn (thick lines).  Thin lines enclose cross section ranges predicted with fitting parameters which are located within 1$\sigma$ around the best fits.

\section{NUCLEOSYNTHESIS MODEL}\label{sec3}

\subsection{Source Energy Spectrum}\label{sec31}

In the PCR and CCR nucleosynthesis model, a total kinetic energy of accelerated CRs is provided by gravitational potential energies of structure and SN explosion energies, respectively.  In the flare nucleosynthesis model, on the other hand, a total kinetic energy of accelerated particles is provided by plasma motions on the stellar surfaces.

The proper source function of CRs or flare particles generated with kinetic energy $E$ at redshift $z_s$ (corresponding to time
$t_s$), i.e., $Q_i(E,z)$, is defined as
\begin{equation}
Q_i(E,z)=D(z) C(z_s)\frac{\phi_i(E,z_s)}{\beta}\delta(t-t_s)
 ~[({\rm GeV/nucleon})^{-1}~{\rm cm}^{-3} {\rm s}^{-1}],
\label{eq4}
\end{equation}
where
$D(z)$ is the ratio of the local number density of H to that of the cosmic average at $z=0$, $C(z_s)$ is an overall amplitude to be fixed with Equation (\ref{eq7}) to a total energy, $\beta$ is the velocity, and $\phi_i(E,z_s)$ is the energy spectrum of a CR or flare particle.  Two types of energy spectra are used.  For the PCR and CCR nucleosynthesis model, the CR injection spectrum of nuclide $i$ is
\begin{equation}
\phi_i(E,z_s)=K_{ip}^{\rm CR}(z_s) [E(E+2E_0)]^{-\gamma/2},
\label{eq5}
\end{equation}
where $K_{ip}^{\rm CR}(z_s)$ is the ratio of number abundance of $i$ to that
of $p$, i.e., $i/p$ of CRs, and $E_0=938$~GeV is the nuclear mass energy per
nucleon.  For the flare nucleosynthesis model, the injection spectrum is
\begin{equation}
\phi_i(E,z_s)=K_{ip}^{\rm flare}(z_s) E^{-s},
\label{eq6}
\end{equation}
where $K_{ip}^{\rm flare}(z_s)$ is the number ratio $i/p$ of flare particle.  In this study, $^4$He abundances in CRs and flares are roughly assumed to be constant at the standard BBN value, i.e., $K_{\alpha p}^{\rm CR}=K_{\alpha p}^{\rm flare}=0.082$.

Amplitudes of the source spectra are set by assuming energy supplies to CRs or flare particles given by
\begin{equation}
{\cal E}(z) =  \int_{E_{\rm min}}^{E_{\rm max}} \sum_i\, A_i Q_i(E,z)\,E~dE,
\label{eq7}
\end{equation}
where
$A_i$ is the mass number or the nucleon number of species $i$.\footnote{There was an error in the normalization of CR energy, i.e., the factor of $A_i$, in \citet{kus2008}.  Abundances of Li, Be, and B calculated by the author should then be multiplied by 0.81.  A similar correction in the normalization seems to be needed for results by Rollinde et al. judging from Equations (1) and (8) of \citet{rol2005}.}
As for the CCR nucleosynthesis, the evolution of CR confinement by a magnetic field is uncertain.  We then assume that the CR confinement is ineffective in the early universe, so that all CRs generated by SNe in structures immediately escape from structures to the intergalactic medium (IGM).  In this case, there is
uniformity of the CR density in the universe.

\subsection{Nuclear Transfer and Reaction Yield}\label{sec32}
We adopt the formulation for CCR nucleosynthesis \citep{mon1977,rol2005,rol2006,kus2008}.  First, we define $N_i(E,z)$ as the number density of a CR species $i$ of energy $E$ at redshift $z$, in units of cm$^{-3}$ (GeV/nucleon)$^{-1}$.  The number abundance relative to that of the background hydrogen $n_{\rm
H}(z)$ is also defined as
\begin{equation}
N_{i,{\rm H}}(E,z)\equiv N_i(E,z)/n_{\rm H}(z). 
\label{eq8}
\end{equation}
We then solve the CR transport equation of $N_{i,{\rm H}}$~\citep{mon1977}
\begin{equation}
\frac{\partial N_{i,{\rm H}}}{\partial t} + \frac{\partial}
{\partial E}(bN_{i,{\rm H}}) + \frac{N_{i,{\rm H}}}{T_{\rm D}} = Q_{i,{\rm H}},
\label{eq9}
\end{equation}
where $b(E,z)\equiv (\partial E/\partial t)$ is the energy loss rate, in units of (GeV/nucleon)~s$^{-1}$, for cosmic expansion or ionization, and $T_{\rm
D}(E,z)$ is the
lifetime against nuclear destruction.  $Q_{i,{\rm H}}(E,z)\equiv Q_i(E,z)/n_{\rm
H}(z)$ is the normalized (comoving) source function.

The expansion loss and ionization loss are expressed in a product of
an energy-dependent term and a redshift-dependent one, $b(E,z)=-B(E)f(z)$. The expansion loss depends on the redshift as $f_{\rm E}(z)=(1+z)^{-1} |dz/dt|H_0^{-1}$, where
$H_0$ is the Hubble constant.  As for the ionization
loss rate we use the fitting formula in \citet{men1971} where we apply the number ratio of $^4$He to H, i.e., He/H=0.08.  The timescale of nuclear destruction of $i$, i.e., $T_{\rm D}=[n_{\rm H}(z)\sigma_{{\rm D},i} \beta]^{-1}$, is estimated with cross sections $\sigma_{{\rm D},i}$ from \citet{ree1974}.

The function $z^\star(E,E',z)$ is defined as in \citet{mon1977},
\begin{equation}
\frac{\partial z^\star}{\partial E}=-\frac{1}{B(E)f(z)}\left|\frac{dz}{dt}\right|\frac {\partial z^\star}{\partial z}.
\label{eq10}
\end{equation}
An energetic nuclide experiences energy loss and time passage (redshift) simultaneously.  Through its transfer, then, the kinetic energy evolves from $E'$ at $z^\star(E,E',z)$ to $E$ at $z$.  No transfer corresponds to the initial redshift of $z^\star(E,E,z)=z$.  CCR particles with energy $E$ at $z$ originate in those with
$E'_s(E,z,z_s)$ at $z_s$.  $E'_s(E,z,z_s)$ satisfies an equation,
$z^\star(E,E'_s,z)=z_s$.  $z^\star(E,E',z)$ is obtained by integrating
Equation~(\ref{eq10}) assuming that the loss process with the greater rate of $b(E,z)$ is dominant all the way from redshift $z^\star$ to $z$.

The transfer equation is solved, and the CCR energy spectrum from a CR generated at redshift $z_s$ is described \citep{kus2008} as
\begin{equation}
\Phi_{i, {\rm H}}(E,z,z_s) = C(z_s) \frac{\phi_i(E'_s,z_s)}{n_{\rm H}^0}\frac{\beta}{\beta'} \frac{|b(E'_s,z_s)|}{|b(E,z_s)|} e^{-\xi(E,E'_s,z)},
\label{eq11}
\end{equation}
where $\Phi_{i,{\rm H}}(E,z,z_s)\equiv \Phi_i(E,z,z_s)/n_{\rm H}(z)$ is
the normalized flux of $i$ per comoving volume with $\Phi_i(E,z,z_s)\equiv
\beta N_i(E,z)_{z_s}$, and $\beta$ and $\beta'$ are the velocities corresponding
to energy $E$ and $E'_s$, respectively.  $n_{\rm H}^0=n_{\rm H}(0)$ is the present
average number density of protons in the universe.  The factor $\xi$ corresponds to the effect of the nuclear destruction through collisions with background protons.  It is given by
\begin{equation}
\xi(E,E'_s,z)=\int_E^{E'_s} \frac{dE''}{\left|b(E'',z^\star(E,E'',z)) T_{\rm D}(E'',z^\star(E,E'',z))\right|}.
\label{eq12}
\end{equation}

The production rate of light nuclide $l$ produced at
redshift $z$ is given by
\begin{equation}
\frac{\partial N_{l,\, {\rm H}}(z,z_s)}{\partial t} =\int \frac{\partial N_{l,\, {\rm H}}(E,z,z_s)}{\partial t} dE = \sum_{i,j} K_{jp}^{\rm tar}(z) \int \sigma_{ij \rightarrow l}^{\rm tot}(E') \Phi_i(E',z,z_s)\, dE',
\label{eq13}
\end{equation}
where 
$\partial N_{l,\, {\rm H}}(E,z,z_s)/\partial t$ is the differential production rate as a function of energy $E$ at production of nuclide $l$,
$K_{jp}^{\rm tar}(z)$ is the target number ratio of nuclide $j$ to
protons in the ISM (PCR model), stellar surface (flare model), and IGM (CCR model); and
$\sigma_{ij \rightarrow l}^{\rm tot}(E')$ is the total cross section
of a reaction $i+j \rightarrow l+X$ with any $X$ that occurs between
a CR nuclide $i$ with energy per nucleon $E'$ and a background species $j$.  Resulting light element
abundances are obtained as the CR-induced production added to BBN yields.  The yields by CR nucleosynthesis are integrations of those for production at $z'$
by CRs generated at $z_s$ over $z'$ and $z_s$, i.e.,
\begin{eqnarray}
\left(\frac{l}{\rm H}\right)_{\rm IGM}(z)&=& \left(\frac{l}{\rm H}\right)_{\rm BBN} \nonumber \\
&&+\int_z^{z_{\rm max}} dz_s \left|\frac{dt}{dz_s}\right| \int _z ^{z_s} dz' \left|\frac{dt}{dz'}\right| \nonumber \\
&&\times \sum_{i,j} K_{jp}^{\rm tar}(z') \int \sigma_{ij \rightarrow l}^{\rm tot}(E') \Phi_i(E',z',z_s)\, dE'.
\label{eq14}
\end{eqnarray}

The target nucleus of reactions we consider is always the $\alpha$ particle, and it is roughly assumed that the $\alpha$ abundances in the ISM and IGM are constant at the standard BBN value, i.e., $K_{\alpha p}^{\rm ISM} =K_{\alpha p}^{\rm IGM} =0.082$.  In addition, we assume that the injection spectrum is constant as a function of time, as an example case.  We can then define the rate of changing abundance ratio of $l$ to H, per unit time of injection, per arbitrary amplitude of source spectrum.  The rate is given by
\begin{eqnarray}
\frac{1}{C(z_s)/n_{\rm H}^0}\frac{\partial}{\partial t_s}\left(\frac{l}{\rm H}\right)_{\rm IGM}(t_s,t)&=&K_{\alpha p}^{\rm tar} \int_{t_s}^t dt' \sum_{i} \int \sigma_{i\alpha \rightarrow l}^{\rm tot}(E') \frac{\Phi_i(E',z',z_s)}{C(z_s)/n_{\rm H}^0}\, dE'.
\label{eq15}
\end{eqnarray}
The total CR (flare particle) energy per hydrogen is proportional to $C(z_s)/n_{\rm H}^0$, and relates to the normalization of resulting yields.

\subsubsection{Cosmic Rays in Structures}\label{sec321}

In astrophysical structures that have been decoupled from the cosmic expansion, energy loss due to the Hubble expansion does not exist.  The rate is then only contributed from the process of ionization loss, and given by
\begin{equation}
 b_{\rm I}(E,z)=-D(z) B_{\rm I}(E).
\label{eq16}
\end{equation}
Using this equation and $T_{\rm D}=[D(z) n_{\rm H}^0\sigma_{{\rm D},i} \beta]^{-1}$, we reduce Equation (\ref{eq11}) to a form
\begin{equation}
 \Phi_{i, {\rm H}}(E,z,z_s) = C(z_s) \frac{\phi_i(E'_s,z_s)}{n_{\rm H}^0}\frac{\beta}{\beta'} \frac{B_{\rm I}(E'_s)}{B_{\rm I}(E)} \exp\left(-\int_E^{E'_s} \frac{n_{\rm H}^0\sigma_{{\rm D},i} \beta''}{B_{\rm I}(E'')}~dE''\right),
\label{eq17}
\end{equation}
where
$\beta''$ is the velocity corresponding to the kinetic energy $E''$.
$^6$He, $^6$Li, and $^7$Li nuclei produced via nucleosynthesis can be processed into $^9$Be.  The probability that a primary product $P$ with initial energy $E$ is converted to $^9$Be is given by
\begin{equation}
 P_{P\rightarrow 9}(E)=\int_{E_{\rm th}}^E~dE'~\frac{n_\alpha(z) \sigma_{P+\alpha\rightarrow ^9{\rm Be}}(E')\beta'}{|b_{\rm I}(E',z)|}=\int_{E_{\rm th}}^E~dE'~\frac{n_\alpha(0) \sigma_{P+\alpha\rightarrow ^9{\rm Be}}(E')\beta'}{B_{\rm I}(E')},
\label{eq18}
\end{equation}
where
$E_{\rm th}$ is the threshold energy in the laboratory frame.
In this equation, the nuclear destruction during the propagation was neglected for the following reason.  Since the source spectra we consider include many low-energy particles, and most of the primary product nuclei tend to have relatively low energies, the effect of nuclear destruction is generally smaller than that of energy loss.
The secondary reactions are assumed to proceed instantaneously in cases of nucleosynthesis at flares and in pregalactic structures.

\subsubsection{Cosmological Cosmic Rays}\label{sec322}

In expanding IGM, the loss rate is contributed also by the expansion loss process.  The loss rate for the flat $\Lambda$CDM model is given by
\begin{equation}
 b_{\rm E}(E,z)=-\left[\Omega_{\rm m}(1+z)^3+1-\Omega_{\rm m}\right]^{1/2} H_0\frac{E(E+2E_0)}{E+E_0}.
\label{eq19}
\end{equation}

The probability that a primary product $P$ ($^6$He, $^6$Li, and $^7$Li) with initial energy $E'$ at $z'$ is converted to $^9$Be by redshift $z$ is given by
\begin{eqnarray}
 P_{P\rightarrow 9}(E',z',z)&=&\int_{E_f(E',z',z)}^E~dE''~\frac{n_\alpha(z^\star) \sigma_{P+\alpha\rightarrow ^9{\rm Be}}(E'')\beta''}{|b(E'',z^\star)|}\nonumber\\
&=&\int_{t'}^{t}~dt''~n_\alpha(z'') \sigma_{P+\alpha\rightarrow ^9{\rm Be}}\left(E''(E',z',z'')\right)\beta''\left(E''(E',z',z'')\right),
\label{eq20}
\end{eqnarray}
where $t'$ is the cosmic time at redshift $z'$, and $z^\star(E',E'',z')$ is obtained by solving Equation (\ref{eq10}).  We note that since the timescale of secondary nuclear reactions is large relative to that of the Hubble expansion, instantaneous thermalization and reaction are not assumed in this case. 

\subsection{Input Parameters}\label{sec33}
We use the code for CCR nucleosynthesis \citep{kus2008}.  Cosmological parameters have been updated.  The standard $\Lambda$CDM model is assumed, and its parameter values are taken from {\it Wilkinson Microwave Anisotropy Probe} seven-year data ($\Lambda$CDM+SZ+lens): $h=H_0/(100~{\rm km}~{\rm s}^{-1}~{\rm Mpc}^{-1})=0.710\pm0.025$, $\Omega_b h^2=0.02258^{+0.00057}_{-0.00056}$, $\Omega_{\rm m}=0.266\pm0.029$, and $\sigma_8=0.801 \pm 0.030$.

The transfer of CR or flare-accelerated $^3$He is newly added in the code.  The cross section of $^3$He destruction via the reaction $^3$He+$p$ is taken from \citet{ree1974}.

\subsubsection{Pregalactic Cosmic Rays}\label{sec331}
We naively assume that the density in pregalactic objects, which merge and later become the Galaxy, is the same as that of the Galaxy.  The density in the Galactic plane is about $2\times 10^{-24}$ g cm$^{-3}$ \citep{pag1997}.  We then take $n_{\rm H}=1$ cm$^{-3}$.  It is assumed that CR $^6$He nuclei decay to $^6$Li before losing energy or experiencing any nuclear reaction since timescales of energy loss and nuclear reaction are much longer than that of $\beta$-decay, i.e., $T_{1/2}=0.8067$~s \citep[][see the next subsection]{ajz1984}.  The CR injection spectrum is assumed to be a power law in momentum (Equation (\ref{eq5})) with spectral index $\gamma=3$ \citep{rol2006,kus2008}.  The lower and upper limits on the range of spectra are taken to be $E_{\rm min}=0.01$~MeV nucleon$^{-1}$ and $E_{\rm max}=10^6$~GeV nucleon$^{-1}$, respectively.  The $^3$He abundance in CRs is assumed to be constant at the standard BBN value of $K_{^3{\rm He}p}^{\rm CR}=1.0\times 10^{-5}$.

\subsubsection{Flare Energetic Particles}\label{sec332}
We assume that the source spectrum of a flare-accelerated energetic particle is a power law in kinetic energy (Equation (\ref{eq6})), and the spectral index is $s=4$ \citep{ram1996,tat2007}.  The range of the spectrum is from $E_{\rm min}=0.01$ MeV nucleon$^{-1}$ to $E_{\rm max}=1$ GeV nucleon$^{-1}$~\citep{tat2007}.  Relative yields of light nuclides do not change when the lower limit is smaller or the upper limit is larger.  This is because low-energy particles have small cross sections due to hindered Coulomb penetration factors, and number densities of high-energy particles are small.  The $^3$He/$^4$He abundance ratio is fixed to be 0.5 \citep{tat2007}:  $K_{^3{\rm He}p}^{\rm flare}=0.5K_{\alpha p}^{\rm flare}$.  The hydrogen number density is taken to be $n_{\rm H}=10^{17}$ cm$^{-3}$.

The $\beta$-decay of $^6$He is neglected since a large portion of energetic $^6$He nuclei thermalize before they $\beta$-decay.  The mean free time of low-energy particles of $E\lesssim {\mathcal O}(10~{\rm MeV})$, for energy loss, i.e., $\tau_{\rm loss}=E/(dE/dt)_{\rm I}$, is shorter than that for nuclear reactions.  The mean free time for a nuclear reaction is
\begin{equation}
\tau_{\rm nuc}=(n_{\rm H}\sigma \beta)^{-1}
    =0.335~{\rm s}\left(\frac{n_{\rm H}}{10^{17}~{\rm cm}^{-3}}\right)^{-1}\left(\frac{\sigma}{10~{\rm mb}}\right)^{-1}\left(\frac{\beta}{0.1}\right)^{-1}.
\label{eq21}
\end{equation}
Because $\tau_{\rm loss}<\tau_\beta(=T_{1/2}/\ln 2)\sim \tau_{\rm nuc}$, the dominant process of $^6$He is the ionization loss.  The effect of $\beta$-decay can then be neglected.

\subsubsection{Cosmological Cosmic Rays}\label{sec333}
We take model 1 of \citet{dai2006}.  The total kinetic energy of CRs is fixed with a parameter $\epsilon$ describing the fraction of SN explosion energy imparted to CRs.  The CR injection spectrum is assumed to be a power law in momentum with spectral index $\gamma=3$, and the energy range is from $E_{\rm min}=0.01$~MeV nucleon$^{-1}$ to $E_{\rm max}=10^6$~GeV nucleon$^{-1}$, similar to the PCR model.  The $^3$He abundance in CRs is taken to be the same as that for the PCR model, $K_{^3{\rm He}p}^{\rm CR}=1.0\times 10^{-5}$.

\section{RESULTS}\label{sec4}

\begin{figure}[htbp]
\begin{center}
\includegraphics[scale=0.4]{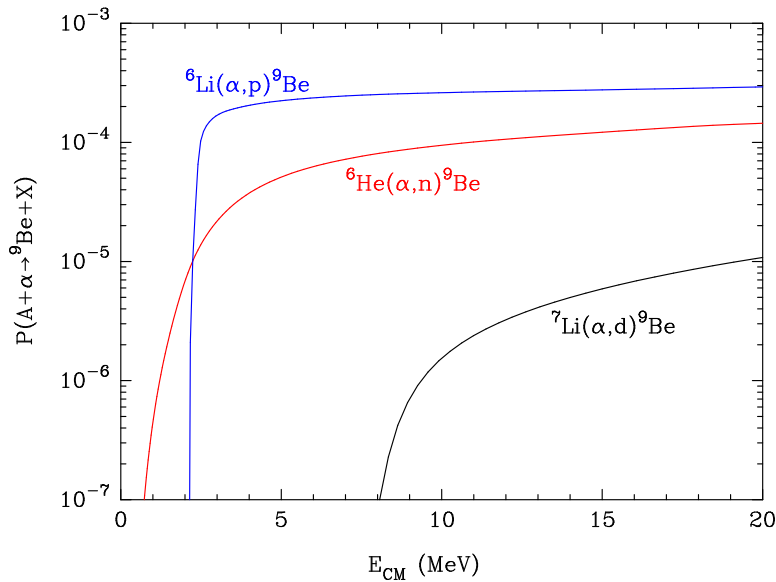}
\end{center}
\caption{Probabilities that nuclides $A$ with an initial center of mass kinetic energy $E_{\rm CM}$ produce $^9$Be through $A$($\alpha$, $X$)$^9$Be reactions: $^6$He($\alpha$, $n$)$^9$Be, $^6$Li($\alpha$, $p$)$^9$Be, and $^7$Li($\alpha$, $d$)$^9$Be.
\label{fig3}}
\end{figure}

Figure \ref{fig3} shows calculated probabilities that nuclides $A$ produce $^9$Be through $A$($\alpha$, $X$)$^9$Be reactions with any $X$ (Equation (\ref{eq18})), as a function of initial energy $E_{\rm CM}$.  The curves are for three reactions, i.e., $^6$He($\alpha$, $n$)$^9$Be, $^6$Li($\alpha$, $p$)$^9$Be, and $^7$Li($\alpha$, $d$)$^9$Be.  We have neglected the effect of cosmic expansion, which is taken into account in the CCR model below.  This calculation is, therefore, applicable to nucleosynthesis in a dense environment such as PCR nucleosynthesis in structures or flare nucleosynthesis on stellar surfaces.

%\placefigure{fig3}

The probabilities increase monotonically as a function of energy since nuclei with higher energy have a greater chance of experiencing nuclear reactions before they lose their energies through ionization loss.  The probability of $^6$Li($\alpha$, $p$)$^9$Be steeply increases at energies right above the threshold, and gently increases after it reaches $2\times 10^{-4}$.  This is because the cross section peaks at 2.4 MeV (see Figure \ref{fig1}), and $^6$Li nuclei with energy $E_{\rm CM}>2.4$~MeV have a relatively small probability of producing $^9$Be until they lose energy down to the peak.

\begin{figure}[htbp]
\begin{center}
\includegraphics[scale=0.4]{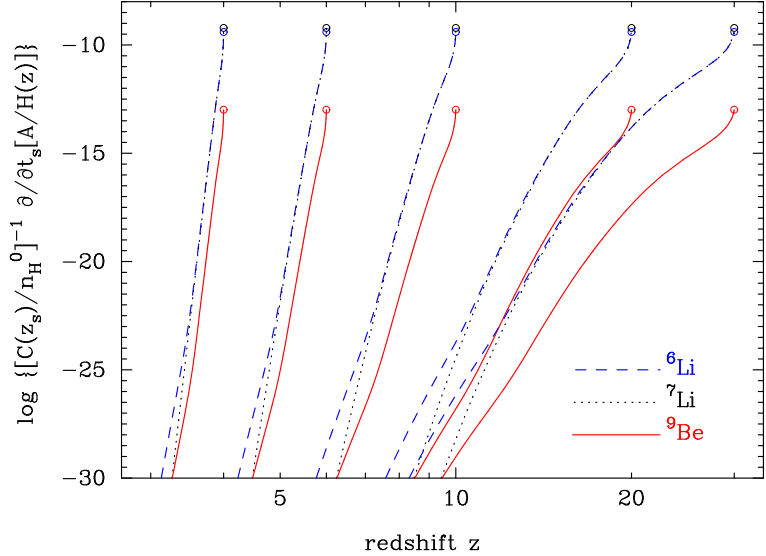}
\end{center}
\caption{Normalized production rates of $^6$Li, $^7$Li, and $^9$Be in the pregalactic cosmic-ray model, in units of (MeV/nucleon)$^{-2}$ s$^{-1}$, as a function of redshift $z$.  Lines are relative rates for production through nuclear reactions operating in the redshift range from 3 to $z(>3)$ by cosmic-rays generated at $z_s=4,~6,~10,~20$, and $30$, respectively.
\label{fig4}}
\end{figure}

Figure \ref{fig4} shows normalized production rates of $^6$Li, $^7$Li, and $^9$Be (Equations (\ref{eq15}), (\ref{eq18})) in the PCR model as a function of redshift $z$.  Curves are for relative production rates through the nuclear reactions occurring at the redshift interval from 3 to $z(>3)$ induced by CRs generated at $z_s=4,~6,~10,~20$, and $30$, respectively.  All $z_s$ cases have the same set of production rates as a function of $t-t_{\rm S}$.  Different lines for the same nuclides then only correspond to different times for injection of CRs.  As for pathways of nucleosynthesis occurring from $z_s$ to $z=3$, $^6$Li is produced through $^4$He($\alpha$, $X$) (99.98 \%) and $^3$He($\alpha$, $p$) (0.02 \%), while $^9$Be is produced through $^4$He($\alpha$, $X$)$^6$Li($\alpha$, $p$) (99.20 \%), $^4$He($\alpha$, $p$)$^7$Li($\alpha$, $d$) (0.79 \%), and $^3$He($\alpha$, $p$)$^6$Li($\alpha$, $p$) (0.01 \%).

%\placefigure{fig4}

\begin{figure}[htbp]
\begin{center}
\includegraphics[scale=0.4]{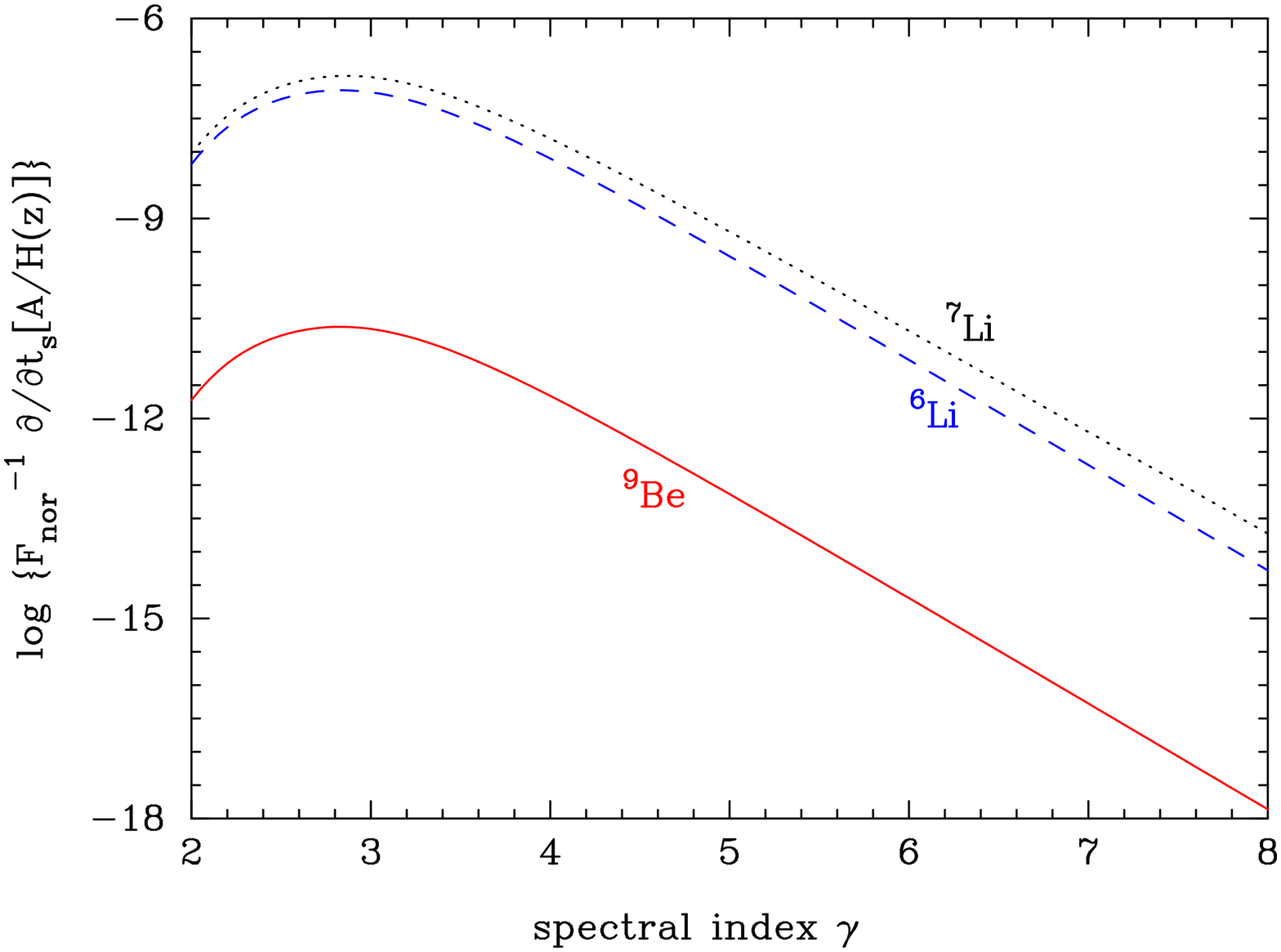}
\end{center}
\caption{Normalized production rates of $^6$Li, $^7$Li, and $^9$Be in the pregalactic cosmic-ray model, in units of MeV$^{-1}$ s$^{-1}$, as a function of the spectral index $\gamma$ for fixed values of $z=3$ and $z_s=4$.  Curves correspond to rates for a fixed injected energy per mass with the normalization factor defined as in Equation (\ref{eqadd1}).
\label{fig5}}
\end{figure}

Figure \ref{fig5} shows normalized production rates of $^6$Li, $^7$Li, and $^9$Be (Equations (\ref{eq15}), (\ref{eq18})) in the PCR model as a function of the spectral index $\gamma$ for fixed values of $z=3$ and $z_s=4$.  In order to estimate production rates for a fixed injected energy per mass, a normalization factor was defined:
\begin{equation}
F_{\rm nor}(\gamma)=\frac{C(z_s)}{n_{\rm H}^0}\int_{E_{\rm min}}^{E_{\rm max}} \frac{\sum_i A_i \phi_i(E,z_s,\gamma)}{\beta}~E\,dE,
\label{eqadd1}
\end{equation}
where the $\gamma$ dependence of spectrum $\phi_i$ was explicitly designated.  The $F_{\rm nor}$ value has a unit of MeV.  The ratio $^6$Li/$^7$Li ranges from 0.7 to 0.28  while the ratio $^9$Be/$^6$Li ranges from $2.9\times 10^{-4}$ to $2.6\times 10^{-4}$ for $2\leq \gamma \leq 8$.  The former ratio is determined by a ratio of $\alpha+\alpha$ cross sections near the threshold.  The latter ratio is relatively constant for the following reason.

%\placefigure{fig5}

As the value of $\gamma$ increases, the contribution of low-energy $\alpha$ particle to the resulting yields of $^{6,7}$Li increases.  Most of $\alpha$+$\alpha$ fusion are, therefore, induced by $\alpha$ particles with energies near thresholds.  The threshold energy of reaction, $\alpha$+$\alpha\rightarrow ^6A+X$, is 22.37 MeV (for $^6A$=$^6$Li) in the CM frame.  The projectile energy in laboratory frame is twice the value.  The kinetic energy of $^6$Li produced by the reaction occurring near threshold is then three eighths of the projectile energy, i.e., 16.78 MeV.  The CM energy for a system of projectile $^6$Li and target $\alpha$ is two fifths of the projectile energy, i.e., $E_{\rm CM,min}=6.71$ MeV.  It is larger than the threshold of the reaction $^6$Li($\alpha$, $p$)$^9$Be, 2.124 MeV.  The $^6$Li nuclei generated by the $\alpha$+$\alpha$ fusion can thus induce $^9$Be production.  Therefore, $^6$Li nuclei produced in the fusion reaction predominantly near its threshold have a finite probability of $^9$Be production: $P(E_{\rm CM})>P(E_{\rm CM,min})=2.4\times10^{-4}$ (Figure \ref{fig3}).

\begin{figure}[htbp]
\begin{center}
\includegraphics[scale=0.4]{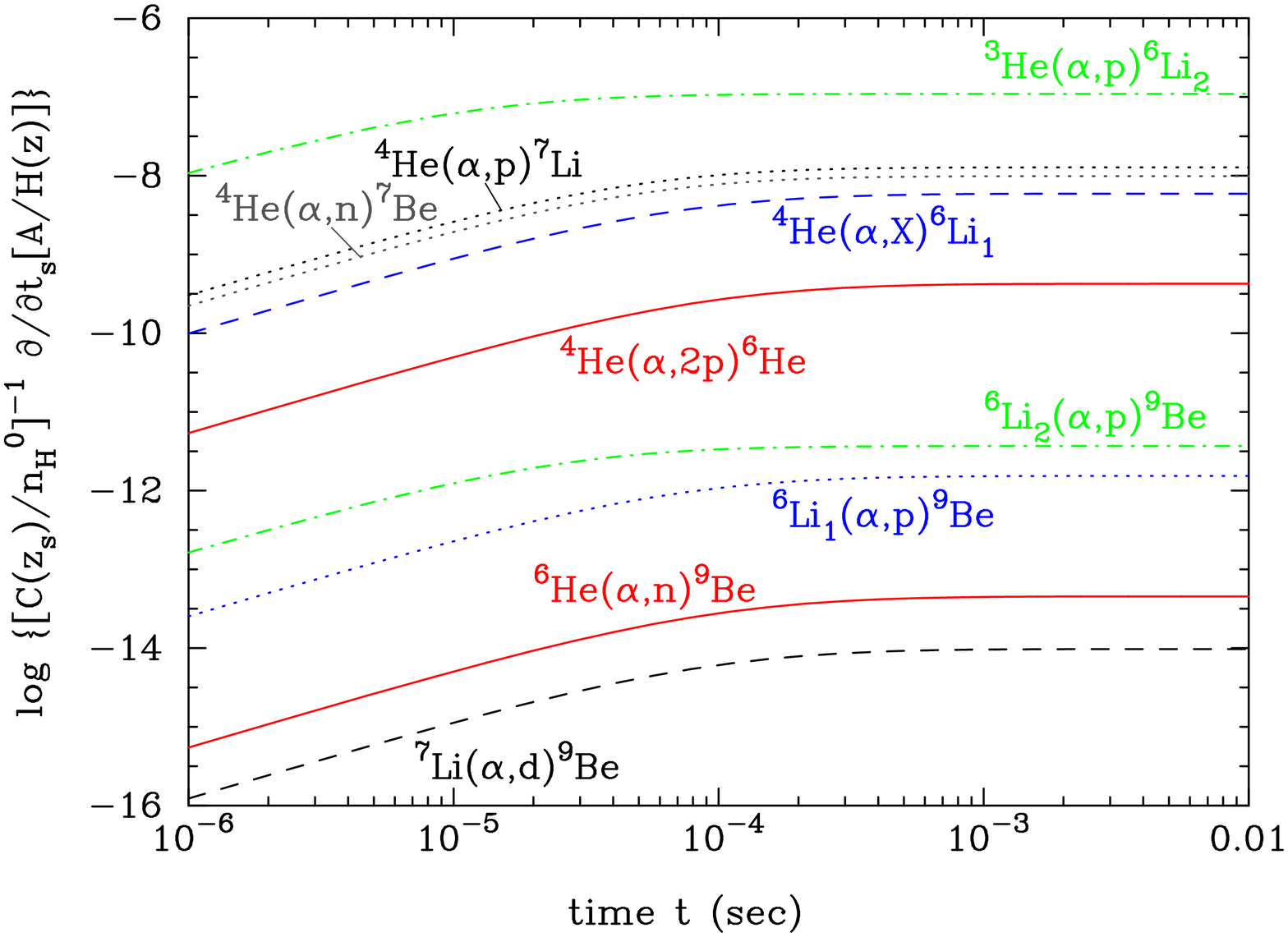}
\end{center}
\caption{Normalized production rates of $^6$Li, $^7$Li, and $^9$Be via specific reactions in the flare model, in units of (MeV/nucleon)$^{-3}$ s$^{-1}$, as a function of time $t$.  Lines are relative rates for production through nuclear reactions by time $t$ induced by flare-accelerated particles generated at $t=0$.
\label{fig6}}
\end{figure}

Figure \ref{fig6} shows normalized production rates of $^6$Li, $^7$Li, and $^9$Be via specific reactions in the flare model as a function of time $t$.  Lines are relative rates for production through nuclear reactions by time $t$ induced by flare-accelerated particles generated at $t=0$.  Since the density in the stellar surface is much higher than that in the IGM, the timescale of nuclear reaction in the stellar surface (human time) is much shorter than that in the IGM (cosmological time).  We then use the time in units of seconds as the parameter for time instead of the redshift.  The result does not depend on the redshift of the particle acceleration $z_s$ which corresponds to the time $t=0$ in this figure.

%\placefigure{fig6}

Compared to the PCR nucleosynthesis model (Figure \ref{fig4}), the $^6$Li yield through $^3$He($\alpha$, $p$)$^6$Li and consequently the $^9$Be yield through $^6$Li($\alpha$, $p$)$^9$Be are relatively larger.  There are two reasons for this.  One is the enhanced abundance ratio of $^3$He.  The other is the difference in nuclear energy spectra.  The softer spectrum of flare particles prefers reactions of smaller threshold energies so that the reaction $^3$He($\alpha$, $p$)$^6$Li ($E_{\rm CM}^{\rm th}=4.019$~MeV) occurs frequently relative to $\alpha+\alpha$ fusion, i.e., $^4$He($\alpha$, $X$)$^{6,7}$A ($E_{\rm CM}^{\rm th}>17$~MeV) in the flare particle model.

\begin{figure}[htbp]
\begin{center}
\includegraphics[scale=0.4]{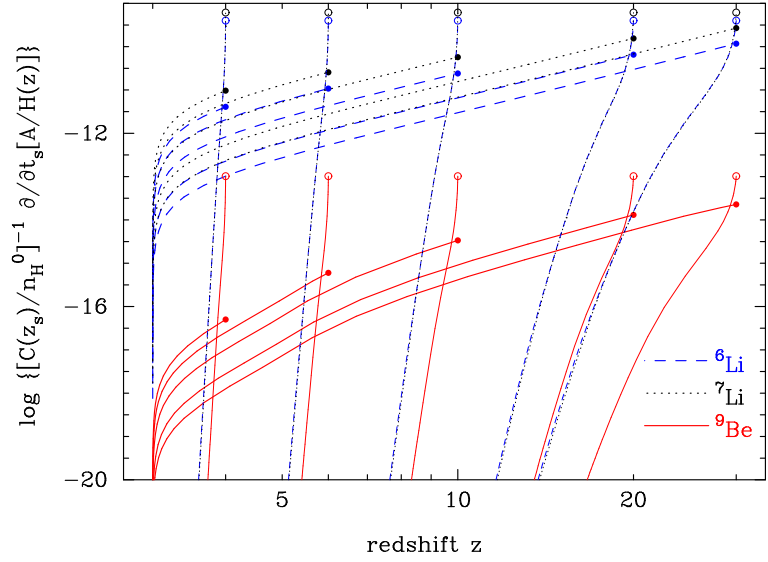}
\end{center}
\caption{Normalized production rates of $^6$Li, $^7$Li, and $^9$Be in the cosmological cosmic-ray model, in units of (MeV/nucleon)$^{-2}$ s$^{-1}$, (thick lines with filled circles attached) as a function of redshift $z$.  Lines are relative rates for production in the redshift range from 3 to $z(>3)$ by cosmic-rays generated at $z_s=4,~6,~10,~20$, and $30$, respectively.  The same quantities in the pregalactic cosmic-ray model (thin lines with open circles attached) are also shown.
\label{fig7}}
\end{figure}

Figure \ref{fig7} shows normalized production rates of $^6$Li, $^7$Li, and $^9$Be in the CCR model (thick lines with filled circles) as a function of redshift $z$ (Equations (\ref{eq15}) and (\ref{eq20})).  Lines are relative rates for production in the redshift range from 3 to $z(>3)$ by CRs generated at $z_s=4,~6,~10,~20$, and $30$, respectively.  The same quantities in the PCR model (thin lines with open circles) are also shown.

%\placefigure{fig7}

Two important differences are found between production rates of CCR and PCR nucleosynthesis models.  First, the effect of cosmological expansion enhances the energy loss rate of CRs, and decreases the production rates.  In the matter-dominated universe of $z\gtrsim 1$, the expansion loss rate is proportional to the Hubble parameter $\propto (1+z)^{3/2}$, while the ionization loss rate is proportional to the matter density $\propto (1+z)^3$.  The effect of cosmic expansion is, therefore, relatively larger in lower redshifts, which is seen as larger differences in the position of open and filled circles for low $z_s$ cases.  Second, the effect of cosmological expansion also enhances the energy loss rate of primary products, i.e., $^6$Li and $^7$Li, and decreases the production rates of secondary product, i.e., $^9$Be, relative to those of primary products.  This effect is also significant at low $z_s$ cases.

The shapes of calculated curves for the CCR and PCR nucleosynthesis models are understood as follows:  in the CCR model, CRs generated at redshift $z_s$ lose their energies through cosmological redshift.  Since the redshift proceeds on a cosmological timescale, the energy loss rate of CRs is relatively small, and CRs can induce nucleosynthesis for a long time.  In the PCR model, on the other hand, since the matter density is much larger than that of the cosmic average value, the energy loss rate of CRs is much larger, and CRs become unable to induce nucleosynthesis more quickly than in the CCR model.  In the PCR model, therefore, most of the $^{6,7}$Li and $^9$Be nuclei are produced right after the generation of CRs at $z_s$.

\begin{figure}[htbp]
\begin{center}
\includegraphics[scale=0.4]{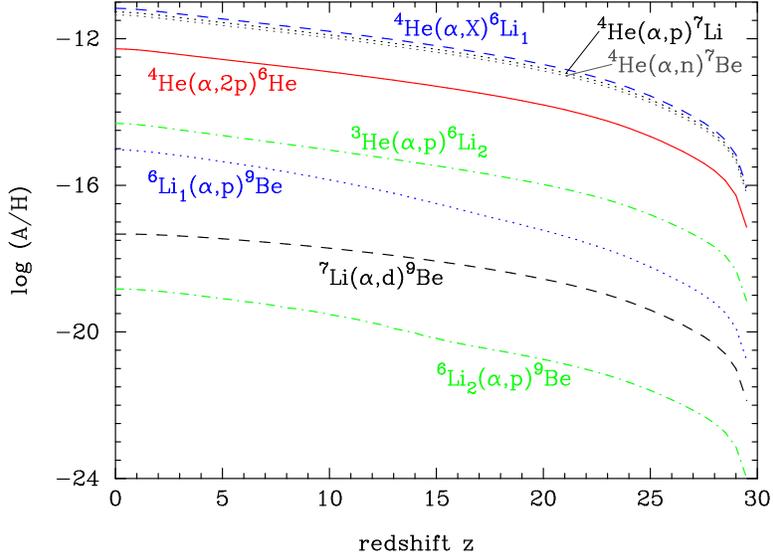}
\end{center}
\caption{Abundances of nuclide $B$ produced through reaction $A$($\alpha$, $X$)$B$ in the cosmological cosmic-ray model using the energy source function of model 1 \citep{dai2006} with 38 \% of SN explosion energy imparted to CRs \citep{kus2008}, as a function of redshift $z$.
\label{fig8}}
\end{figure}

Figure \ref{fig8} shows resulting abundances of nuclide $B$ produced through reactions $A$($\alpha$, $X$)$B$ in the CCR model using the energy source function of model 1 of \citet{dai2006} with 31 \%/0.81=38 \%\footnote{The error in the normalization of CR energy in \citet{kus2008} was corrected (see footnote in Section \ref{sec31}).} of the SN explosion energy imparted to CRs \citep{kus2008}, as a function of redshift $z$.

%\placefigure{fig8}

We note that the total yield of $^6$He+$^6$Li is $\sim$20 \% smaller at $z=3$ than in \citet{kus2008}.  This difference is caused by an uncertainty in the reaction cross section of the $\alpha$+$\alpha$ fusion reaction.  The total cross section of $^4$He($\alpha$, $X$)$^6A$ adopted in this study, i.e., \citet[][Graph IV, the sum of the two lines in the bottom panel]{rea1984}, is smaller than that adopted in \citet{kus2008}, i.e., \citet[][Graph IV, upper panel]{rea1984}.

\section{PREDICTED $^9$Be/$^6$Li RATIOS}\label{sec5}

\subsection{Pregalactic Cosmic Rays}\label{sec51}
Although structure formations would proceed with metal enhancements by SNe, the previous study of PCR energized by structure formation shocks~\citep{suz2002} did not include time evolution of CNO nuclear abundances and $^9$Be production from CNO spallation.  It combined the standard GCR nucleosynthesis and PCR nucleosynthesis including only the $\alpha$ particle as a component of CR species.  The abundance of $^9$Be produced via the CNO spallation scales as CNO abundances, i.e., [CNO/H], in the ISM in the structure formation epoch.  Since the value [CNO/H] increases roughly as a total number of SN events in the standard GCR nucleosynthesis model, it is very small in the structure formation epoch at a low metallicity.  The produced abundance of $^9$Be would then be very small.  The neglect of the CNO spallation in their model is, therefore, reasonable.
The yield ratio of $^9$Be/$^6$Li in the PCR model is
\begin{equation}
\left.\frac{^9{\rm Be}}{^6{\rm Li}}\right|_{\rm spa}=0.
\label{eq22}
\end{equation}
The $^9$Be production through $^6$He, $^6$Li, and $^7$Li considered in the present study does not depend on metallicity.  The calculated ratio is read from circles in Figure \ref{fig4}, and given by
\begin{equation}
\left.\frac{^9{\rm Be}}{^6{\rm Li}}\right|_{\rm fus}=2.8\times 10^{-4}.
%2.8163e-4
\label{eq23}
\end{equation}
This value is interestingly high.  After MPSs are formed of material with this abundance ratio, the ratio in MPSs decreases during the life of the star until we observe it since $^6$Li is burned more easily in a star than $^9$Be is~\citep{pag1997}.  Suppose that the $\alpha+\alpha$ fusion is really the cause of an observed high $^6$Li abundance, and $^6$Li abundance has been depleted by 0.8 dex as assumed in \citet{tat2007}.  The $^9$Be abundance is then predicted to be $^9$Be/H=$10^{0.8}$($^6$Li/H)$_{\rm obs}\times ^9$Be/$^6$Li$|_{\rm fus}\sim 1.1\times 10^{-14}$.  This level is almost excluded from the present observational upper limit~\citep{ito2009}.  Therefore, an elevated abundance may be detected in high precision observations in the near future.

This ratio of $^9$Be/$^6$Li asymptotically approaches $2.4\times 10^{-4}$ when the spectral index $\gamma$ increases (Figure \ref{fig5}).  Note that the ratio is independent of the $^4$He abundance of projectile.  For example, a high $\gamma$ value is expected for outer mass shells in Type Ic SNe \citep{fie2002,nak2004}:  the value $\gamma=7.2$ is derived when we take the nonrelativistic limit, and assume that the flux of particles with momenta from $p$ to $p+dp$ is given by $\phi(p)dp\propto d[M(>v)v]/dv$, with $M(>v)\propto v^{-7.2}$ being the integrated mass distribution as a function of velocity.  This $\gamma$ value corresponds to $^9$Be/$^6$Li$=2.6\times 10^{-4}$ in the PCR model.  The yield ratio for $\alpha$-fusion reactions in the Type Ic SN model is $^9$Be/$^6$Li$=1.8\times 10^{-3}$ taking into account a difference in nuclear abundances in the CSM and ISM.  This value is estimated as follows:  in Equation (\ref{eq18}), the ionization loss rate is proportional to the electron number density, $\sum_i Z_i n_i$ with $Z_i$ and $n_i$ the proton number and the number density of nuclear species $i$.  The probability of $^9$Be production is then proportional to $n_\alpha(z)/\sum_i Z_i n_i(z)$.  The above ratio is thus derived considering that the CSM in Type Ic SNe is composed predominantly of $\alpha$ particles only~\citep{nak2006}.

\subsection{Flare Particles}\label{sec52}
\citet{tat2007} have calculated yield ratios for nucleosynthesis taking into account CNO spallation producing $^9$Be.  They have shown calculated values for spectral index $s=4$, i.e.,
\begin{equation}
\left.\frac{^9{\rm Be}}{^6{\rm Li}}\right|_{\rm spa}=3.5\times 10^{-6}\frac{Z/Z_\sun}{10^{-3}}.
\label{eq24}
\end{equation}
This ratio depends on stellar metallicities which are observables.  $^9$Be production through $^6$He, $^6$Li, and $^7$Li, on the other hand, does not depend on metallicity.  The ratio calculated in the present study (see Figure \ref{fig6}) is
\begin{equation}
\left.\frac{^9{\rm Be}}{^6{\rm Li}}\right|_{\rm fus}=4.6\times 10^{-5}.
%4.5663e-05
\label{eq25}
\end{equation}
In the present setting of flare particle abundances and the index of the source spectrum, the two-step $\alpha$-fusion reaction is the dominant process producing $^9$Be on surfaces of MPSs of $Z\lesssim 1.3\times 10^{-2}Z_\sun$.

\subsection{Cosmological Cosmic Rays}\label{sec53}
A yield ratio in the CCR model taking into account the primary process of CO spallation~\citep{kus2008} calculated for model 1 of \citet{dai2006} at $z=3$ is given by
\begin{equation}
\left.\frac{^9{\rm Be}}{^6{\rm Li}}\right|_{\rm spa}=0.015.
\label{eq26}
\end{equation}
This value seems somewhat high.  For example, even if we assume no reduction of $^6$Li abundance in the stellar surface, a $^6$Li production up to the possible plateau level, i.e., $^6$Li/H=$6\times 10^{-12}$~\citep{asp2006}, is accompanied by a $^9$Be production of $^9$Be/H$\sim 10^{-13}$.  This $^9$Be/H abundance is about 10 times higher than the observational upper limit on primordial abundance~\citep{ito2009}.  This yield ratio, however, depends on the time evolution of CR energy content and intergalactic metallicity in the early universe which are largely uncertain.  If the metallicity of the IGM is lower in our universe than supposed in model 1, the $^9$Be yield and $^9$Be/$^6$Li ratio are smaller.  In addition, as observed in Section~\ref{sec4}, CRs accelerated at lower redshifts $z_s$ produce smaller amounts of $^9$Be relative to $^6$Li.  If the CR energies transferred from SN kinetic energies are relatively larger at lower redshifts than in model 1, the $^9$Be/$^6$Li ratio is smaller.

Ratios between rates of $^9$Be production only through $^6$He, $^6$Li, and $^7$Li and those of $^6$Li by $\alpha+\alpha$ fusion are calculated in this paper (Figure \ref{fig7}).  They depend on $z_s$, and their range is $1.3\times 10^{-5} \leq ^9$Be/$^6$Li$\leq 2.2\times 10^{-4}$.  
%1.3298e-5, 6.01602e-5, 1.5097e-4, 2.1611e-4, 2.0975e-4
The yield ratio at $z=3$ is calculated by integrating production rates with a weight of CR energy amplitude over redshift $z_s$, and is found (Figure \ref{fig8}) to be
\begin{equation}
\left.\frac{^9{\rm Be}}{^6{\rm Li}}\right|_{\rm fus}=1.4\times 10^{-4}.
%1.392530e-4
\label{eq27}
\end{equation}
This ratio is determined with only primordial abundances of $^4$He and $^3$He, and independent of information on a metal pollution of the universe.  The derived ratio is, therefore, a lower limit on the initial abundance ratio of material for observed MPSs.  A contribution of $^9$Be through CNO spallation adds to this ratio.

\subsection{$^9$Be/H versus Fe/H Plot}\label{sec54}
\begin{figure}[htbp]
\begin{center}
\includegraphics[scale=0.4]{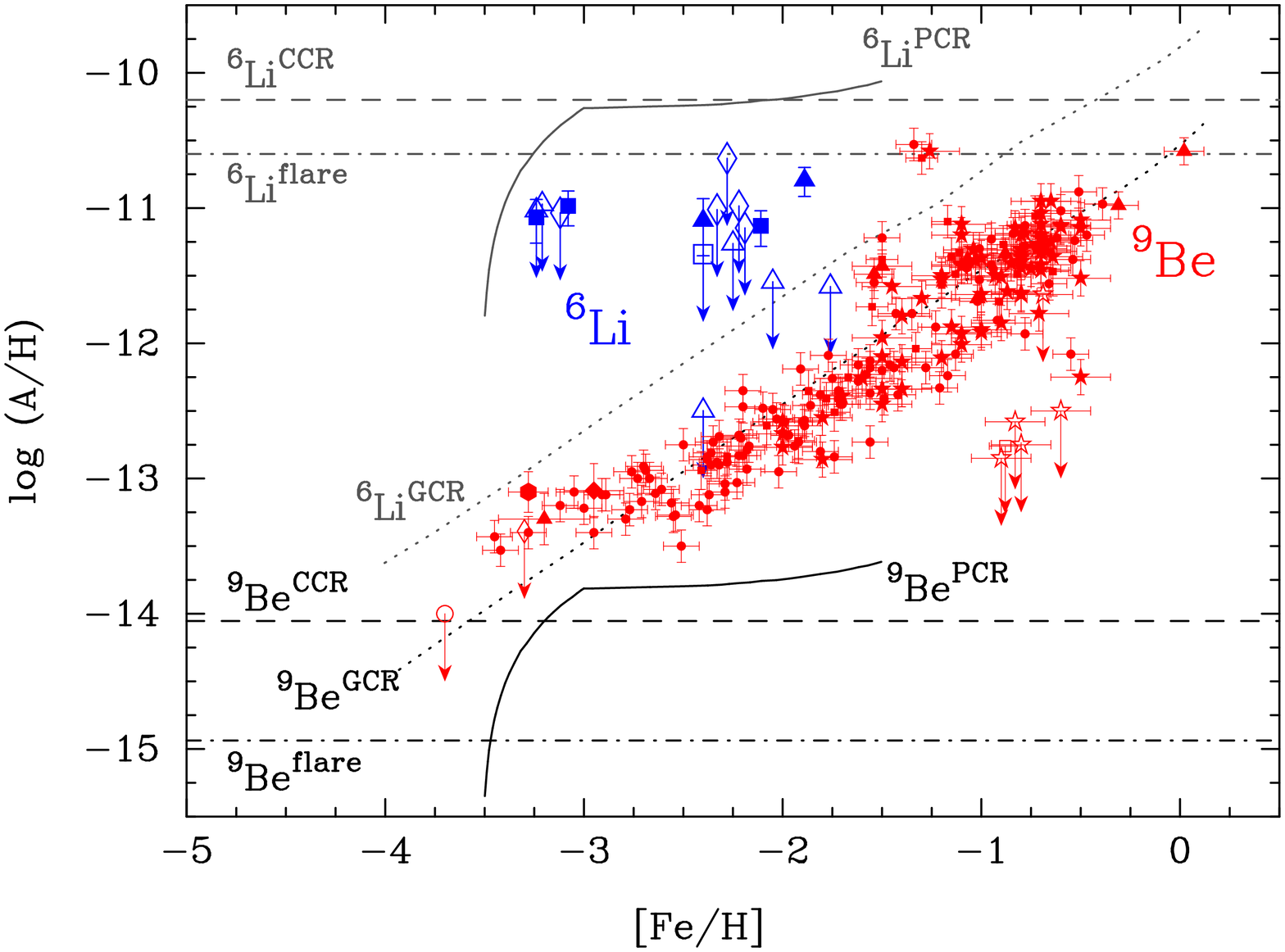}
\end{center}
\caption{Abundances of $^6$Li and $^9$Be produced in models of nucleosynthesis by pregalactic cosmic-ray (PCR: solid lines), cosmological cosmic-ray (CCR: dashed), and flare-accelerated nuclides (dot-dashed), as a function of iron abundance.  Results for the standard Galactic cosmic-ray nucleosynthesis model \citep{pra2012} are shown with dotted lines.  $^6$Li data are from \citet[][open and filled large squares for upper limits and detections at 2 $\sigma$, respectively]{asp2008}, \citet[][open large diamonds for upper limits]{gar2009}, and \citet[][open and filled large triangles for upper limits and detections]{ste2012}.  $^9$Be data are from \citet[][open and filled small diamonds for upper limits and detections]{pri2000b}, \citet[][filled hexagons]{pri2000a}, \citet[][filled small triangles]{boe2006}, \citet[][open circle]{ito2009}, \citet[][open and filled small squares]{tan2009}, \citet[][open and filled stars]{smi2009}, and \citet[][filled circles]{boe2011}.
\label{fig9}}
\end{figure}

Figure \ref{fig9} shows abundances of $^6$Li and $^9$Be originating from $\alpha+\alpha$ fusion reactions in models of PCR (solid lines), CCR (dashed), and flare (dot-dashed) nucleosynthesis, as a function of iron abundance.  We assumed the following settings as examples for drawing these lines.  PCR: $^6$Li abundance in Model I of \citet{suz2002} was scaled to fit the observed abundance ($^6$Li/H$\sim 10^{-11}$) to be seen after a possible depletion in stellar surface during pre-main- and main-sequence evolution by a factor of $10^{0.8}$~\citep{tat2007};  $^9$Be abundance was estimated by multiplying the yield ratio (Equation (\ref{eq23})) to the $^6$Li abundance.  CCR: $^6$Li abundance was arbitrarily chosen to fit the observed level and the depletion factor as in the PCR case.  $^9$Be abundance was estimated by multiplying the yield ratio (Equation (\ref{eq27})) by the $^6$Li abundance.  Flare: $^6$Li abundance was arbitrarily chosen to fit the observed level after a depletion during only main-sequence evolution by a factor of $10^{0.4}$~\citep{tat2007};  $^9$Be abundance was estimated by multiplying the yield ratio (Equation (\ref{eq25})) to the $^6$Li abundance.  

%\placefigure{fig9}

Observed abundances of $^6$Li and $^9$Be in MPSs are shown with various markers (see the figure caption for references).  We plot only $^6$Li data derived from recent analyses of the lithium line profile taking into account effects of convection~\citep{cay2007}.  Dotted lines show a result in a standard GCR nucleosynthesis model \citep{pra2012}.  It is seen that stellar $^9$Be abundances can be well fitted by the GCR model, while those of $^6$Li deviate significantly from the model prediction.

In Model I of \citet{suz2002}, an injection of CR from structure formation shocks is assumed to occur from $t_{\rm SF}=0.2$ Gyr during $\tau_{\rm SF}=0.1$ Gyr with its spectral index fixed to be $\gamma=3$.  Although the model successfully explains enhanced $^6$Li abundances in stars of [Fe/H]$\geq -3$, it cannot predict the high $^6$Li abundance in stars of lower metallicities.  If the PCR nucleosynthesis produces  $^6$Li at the abundance level of $^6$Li/H$\sim 10^{-10.2}$ at [Fe/H]$\lesssim -3.3$, the abundance of coproduced $^9$Be exceeds that originating from standard GCR nucleosynthesis.  

Lastly, we note that all CCR, PCR, and GCR nucleosynthesis must contribute to abundances of light elements, i.e., $^6$Li, $^9$Be, and B, in MPSs although we do not know their relative importance precisely yet.  In this paper, we separately deal with the three kinds of nucleosynthesis.  In the universe, however, CCR first enhances light element abundances in the IGM , which later falls in the early Galaxy.  The formation of the Galaxy then proceeds with formation of structures associated with CR injections.  Stellar activities in such structures gradually pollute the structures.  Finally, the Galaxy formation completes, and the standard GCR nucleosynthesis model would describe the abundance evolution of light elements well.  The three CR nucleosynthesis models will be unified with the aid of studies on cosmic and Galactic chemical evolution in the future.

\section{CONCLUSIONS}\label{sec6}

We have suggested the possibility of $^9$Be production by two-step $\alpha$-fusion reactions of CR or flare-accelerated $^{3,4}$He via intermediate nuclides $^6$He and $^{6,7}$Li which occur in the IGM, pregalactic structure, and stellar surfaces.  We calculated probabilities that nuclides $^6$He and $^{6,7}$Li with variable kinetic energy at production synthesize $^9$Be through reactions with an $\alpha$ particle.  Production rates of $^{6,7}$Li and $^9$Be are then calculated in three models, i.e., nucleosynthesis by pregalactic and cosmological CRs and flare energetic particles, by taking into account transfers and nuclear reactions of CR or flare particles.

What we have found is as follows.  

1) In the pregalactic CR nucleosynthesis model, $^9$Be is produced mainly via \\
$^4$He($\alpha$, $X$)$^6$Li($\alpha$, $p$)$^9$Be.  $^6$Li has the highest probability of producing $^9$Be among all primary product nuclides.  This stems from the highest cross section of the reaction $^6$Li($\alpha$, $p$)$^9$Be.  The calculated result of the $^9$Be/$^6$Li ratio indicates that a $^6$Li production up to the possible high level observed in MPSs is accompanied by production of $^9$Be at a level of the present observational upper limit.  An enhanced abundance of $^9$Be may, therefore, be observed in MPSs in the future if this pregalactic CR nucleosynthesis is the cause of the observed high abundance of $^6$Li.

2) In the flare nucleosynthesis model, $^9$Be is produced mainly via $^3$He($\alpha$, $p$)$^6$Li($\alpha$, $p$)$^9$Be.  The original seed of $^9$Be is $^3$He since the $^3$He/$^4$He ratio is assumed to be high (as in the Sun), and the softer energy spectrum prefers the reaction $^3$He($\alpha$, $p$)$^6$Li with a lower threshold than $^4$He($\alpha$, $X$)$^6$Li.  Since the two-step $\alpha$-fusion reactions need rather high initial energies of seed $^{3,4}$He nuclei, the softer source spectrum in the model results in a low yield of $^9$Be relative to that of $^6$Li.  The calculated ratio $^9$Be/$^6$Li is then $\sim 0.2$ times as large as that of the pregalactic CR model.  The $^9$Be yield in the present process is metallicity-independent lower limits on abundances in MPSs, and is larger than the yield through CNO spallation in MPSs with metallicity $Z\lesssim 1.3\times 10^{-2}Z_\sun$.

3) In the cosmological CR model, $^9$Be is produced mainly via $^4$He($\alpha$, $X$)$^6$Li($\alpha$, $p$)$^9$Be.  Since the cosmic expansion enhances the energy loss rate of CRs, yields of nucleosynthesis decrease from those in the pregalactic CR model.  The decreases in yields of the secondary product $^9$Be are larger than those of the primary products $^{6,7}$Li since the former results from two-step energy loss processes while the latter experience energy loss only once during the CR propagation.  The calculated ratio $^9$Be/$^6$Li is then $\sim 0.05$--$0.8$ times as large as that of the pregalactic CR model for the propagation of CRs from redshift $4\leq z_s\leq 30$ to 3.  When a model of CR energy profile in the universe as a function of redshift \citep{dai2006} is adopted, the integrated ratio $^9$Be/$^6$Li is $\sim 0.5$ times as large as that of the pregalactic CR model.  This $^9$Be yield is a metallicity-independent lower limit.  To this yield, a contribution of CNO spallation which depends on the time evolution of metallicity in the cosmic chemical evolution model should be added.

\acknowledgments
This work has been supported by Grants-in-Aid for the Japan Society for
the Promotion of Science (JSPS) Fellows (21.6817), and for Scientific Research from the Ministry of Education, Science, Sports, and Culture (MEXT), Japan,
No.22540267 and No.21111006 (Kawasaki), and by World Premier International Research Center Initiative (WPI Initiative), MEXT, Japan.

%\clearpage

%\clearpage

%\clearpage

\end{document}